\documentclass[11pt]{article}

\usepackage[final]{acl}

\usepackage{times}
\usepackage{latexsym}
\usepackage[T1]{fontenc}
\usepackage[utf8]{inputenc}
\usepackage{microtype}
\usepackage{inconsolata}
\usepackage{graphicx}
\usepackage{booktabs}
\usepackage{amsmath}
\usepackage{amsfonts}
\usepackage{multirow}
\usepackage{tcolorbox}
\usepackage{algpseudocode}

\title{Agentic Economic Modeling}

\author{Bohan Zhang \\
  University of Michigan \\
  {\normalsize \texttt{zbohan@umich.edu}}
  \And
  Jiaxuan Li\\
  Amazon\\
  {\normalsize \texttt{jiaxuan@amazon.com}}
  \And
  Ali Horta\c{c}su\\
  University of Chicago\\
  {\normalsize \texttt{hortacsu@uchicago.edu}}
  \And
  Xiaoyang Ye\\
  Amazon\\
  {\normalsize \texttt{yexiaoys@amazon.com}}
  \AND
  Victor Chernozhukov\\
  MIT\\
  {\normalsize \texttt{vchern@mit.edu}}
  \And
  Anqi Ni\\
  Amazon\\
  {\normalsize \texttt{nianq@amazon.com}}
  \And
  Edward W Huang\\
  Amazon\\
  {\normalsize \texttt{ewhuang@amazon.com}}
  }
\author{
\bf Bohan Zhang$^{1}$, Jiaxuan Li$^{2}$, Ali Horta\c{c}su$^{3}$, Xiaoyang Ye$^{2}$,\\
\bf Victor Chernozhukov$^{4}$, Anqi Ni$^{2}$, {\normalfont and} Edward W Huang$^{2}$ \\
\textsuperscript{1}University of Michigan,
\textsuperscript{2}Amazon, \\
\textsuperscript{3}University of Chicago, \textsuperscript{4}Massachusetts Institute of Technology \\
\textsuperscript{1}\texttt{zbohan@umich.edu}, 
\textsuperscript{2}\texttt{\{jiaxuan,yexiaoys,nianq,ewhuang\}@amazon.com} \\
\textsuperscript{3}\texttt{hortacsu@uchicago.edu}, \textsuperscript{4}\texttt{vchern@mit.edu}
}
\begin{document}
\maketitle

\begin{abstract}
We introduce Agentic Economic Modeling (AEM), a framework that aligns synthetic LLM choices with small-sample human evidence for econometric inference. AEM first generates task-conditioned synthetic choices via LLMs, then learns a bias-correction mapping from task features and raw LLM choices to human-aligned choices, upon which standard econometric estimators perform inference to recover demand elasticities and treatment effects. We validate AEM in two experiments. In a large scale conjoint study, using only 10\% of the original data to fit the correction model lowers the error of the demand-parameter estimates, while uncorrected LLM choices increase the errors. In a regional field experiment, a mixture model calibrated on 10\% of geographic regions estimates a treatment effect of -65$\pm$10 bps on the hold-out regions, closely matching the full human experiment (-60$\pm$8 bps). These results demonstrate AEM's potential to improve RCT efficiency and represent a step toward LLM-based counterfactual generation.
\end{abstract}

\section{Introduction and Related Work}

At its core, making business decisions requires understanding cause and effect. For example, when an e-commerce company considers applying new features, we need to understand counterfactuals: how will new features cause changes in customers' purchasing decisions? The fundamental challenge is that we can never directly observe both what happens when we make a change and what would have happened if we hadn't. Traditionally, this required either running extensive experiments, conducting surveys, or applying complicated causal inference techniques to historical data.

Large Language Models (LLMs) offer a promising solution through their ability to generate synthetic responses based on vast amounts of world knowledge. The ``homo silicus'' paradigm~\citep{hortone2024} treats LLMs as simulated economic agents that can be placed in decision environments to generate choice behavior at scale. This provides a low-cost way to explore experimental designs and behavioral patterns before large-scale human data collection, while recognizing that such simulations remain imperfect and require empirical validation.~\citet{binz2025centaur} further shows that fine-tuned LLMs can predict human behavior across diverse cognitive tasks. Recent work further explores how theory-guided LLM personas can generalize across strategic environments~\cite{manning2025general}. For economists, this opens an attractive possibility~\citep{Korinek2023GenerativeAF,NBERw33033}: replacing expensive experiments with large-scale synthetic choice data while still recovering demand elasticities and treatment effects.

However, a fundamental challenge remains: LLM-generated choices often diverge systematically from real customer behavior. Off-the-shelf LLMs exhibit systematic preference bias and lower heterogeneity~\citep{chen2023emergenceeconomicrationalitygpt,delriochanona2025generativeaiagentsbehave}. Models display machine bias with low within-topic variance~\citep{boelaert2025machine}, and alignment choices shape results systematically~\citep{lyman2025balancing,kozlowski2025simulating}. As a result, naive inference on synthetic LLM outputs without bias correction is unreliable~\citep{wang2024large,dubois2024alpacafarmsimulationframeworkmethods,li2025llm}.

Our approach builds on the data-augmentation line from  \citet{wang2024large}, which formalizes bias correction for LLM-generated choice data via learned mappings between human and LLM responses. A related approach,  prediction-powered inference (PPI) ~\citep{angelopoulos2023ppi,angelopoulos2023ppiplus},  uses the predictor on the large unlabeled sample to estimate the predictable part of the outcome, and uses the labeled sample only to correct bias through residuals. This yields unbiasedness regardless of predictor quality, provided the predictor is treated as fixed relative to the labeled sample used to compute residuals. \citet{ji2025predictions} extend PPI to show that AI predictions can serve as valid surrogates when properly calibrated. \citet{ludwig2025llm} provide a foundational econometric framework distinguishing prediction problems (where LLMs excel) from estimation problems (where human calibration is essential). \citet{ruan2025calm} demonstrate that language models can boost RCT power without statistical bias when properly integrated.

In this paper, we introduce Agentic Economic Modeling (AEM), a framework that aligns synthetic LLM choices with small-sample human evidence for reliable econometric inference. The key innovation builds upon three ideas: First, we recognize that LLMs, having absorbed vast world knowledge, can help us better predict customer behaviors. Rather than using raw LLM predictions directly, we transform their knowledge-rich but biased outputs into estimates of human decision-making. This is inspired by recent work~\cite{wang2024large} that formalizes statistical augmentation through learned mappings between human and LLM responses, and shows that naive substitution can worsen bias. Our work follows this general insight and extends it from conjoint tasks to a broader Generation–Correction–Inference pipeline and to large-scale RCT settings with aggregate outcomes. Second, drawing on the random coefficient discrete choice literature~\citep{dc-rum}, we recognize that community-level choices are better represented by mixtures of individual agent decisions rather than a single agent model. Mirroring the heterogeneity of real communities, we generate predictions from multiple AI personas, each reflecting a distinct decision-making style. \citet{manning2025general} demonstrate that persona mixtures can capture population-level heterogeneity, while Mixture-of-Personas~\citep{bui2025mixture} formalizes this as a probabilistic framework. Third, we leverage this mixture model insight to correct for LLM biases. Rather than trying to make individual LLMs perfectly mimic human behavior, we generate diverse synthetic agents through different personas, and learn the optimal weights to match observed choice patterns. This allows us to span the space of possible decision rules while anchoring predictions to reality using minimal real-world data.

We validate our framework in two RCT settings. First, in a large-scale conjoint experiment (millions of observations), bias correction substantially improves point-estimate accuracy using just 10\% of the test product's human data, validating prior work~\citep{wang2024large}. Then, in a regional field experiment, our mixture model estimated a treatment effect of -65 bps ($\pm$ 10 bps) on the hold-out regions, closely matching the actual human experiment effect (-60 $\pm$ 8 bps). These results demonstrate AEM's potential to improve RCT efficiency and represent a step toward LLM-based counterfactual generation.

\section{Connection to Random Coefficient Discrete Choice Modeling}\label{sec:mix-model}

In economic settings, decisions are made by intelligent agents following optimization principles and/or heuristics. A frequently encountered example is from the discrete choice/random utility model (DC-RUM)~\citep{dc-rum}. The agent is commonly modeled as making a choice based on a softmax model:
\begin{equation}
\Pr(i \text{ chooses alternative } j)= \frac{\exp(w_i^{T} X_j)}{\sum_k \exp(w_i^{T} X_k)}      
\end{equation}

Agent $i$ here is characterized by the vector of weights $w_i$ and $X_j$ is the feature vector of option $j$. To explain the behavior of a community of such agents, a better representation is provided by the \textit{mixture} of choices by individual agents:
\begin{equation}
\resizebox{0.95\columnwidth}{!}{$
\Pr(\text{community chooses } j)=
\sum_{i \in \mathcal{I}} \theta_i
\frac{\exp(w_i^{T} X_j)}
{\sum_k \exp(w_i^{T}X_k)}
$}
\end{equation}
where $\theta_i$ represents the weight given to the choices of agent $i$. This framework is frequently called the ``random coefficient discrete choice'' framework, where the main idea is to specify a discrete support for the distribution of agent types, and apply discrete probability weights to represent the choices of the community as a mixture probability generated by these weights.

\begin{table*}[htbp]
  \centering
  \resizebox{0.8\textwidth}{!}{
  \begin{tabular}{c|c|c}
    \toprule
    \textbf{Stage} & \textbf{Input $\to$ Output} & \textbf{Description} \\
    \midrule
    Generation & task $x$ $\to$ choices $z$ & LLM produces synthetic decisions for diverse personas. \\
    Correction  & $(x,z)$, labels $y$ $\to$ $\hat y$ & Learn bias-correction so LLM aligns with humans. \\
    Inference    & $(x,\hat y)$ $\to$ $\theta$ & Econometric models derive elasticities, effects. \\
    \bottomrule
  \end{tabular}}
  \caption{Summary of the Generation–Correction–Inference pipeline.}
  \label{tab:gci_pipeline}
\end{table*}

Where does the ``agentic'' framework come into play here? We can think of pre-generating a set of $N$ agents, each of whom have a different decision rule, characterized in the above model as $w_i$. Each pre-generated agent is going to make a choice that is logically consistent with its parameters, but the aim is to calibrate, or fine-tune the population of agents by learning the weights $\theta_i$ that best fit the observed choice probabilities in the real world data.

The bigger idea, however, is that this framework is not limited to the very simple single layer perceptron of McFadden, and allows for a plethora of decision rules or agent ``personas'' that are consistent with what a large foundational LLM has learned from its immense input set. Whatever the underlying ``agentic'' decision maker is, our framework calls for a calibration or fine-tuning stage where we learn the weights $\theta_i$ for each of the underlying agent personas:
\begin{multline}
\Pr(\text{community chooses } j) \\ = \sum_{i \in \mathcal{I}} \theta_i \Pr(\text{agent persona } i \text{ chooses } j | X)  
\end{multline}

The idea here is that to model the observed set of choices of a community of customers, a single agent persona's decision framework will typically not be enough to capture the richness of the decision-rule space. However, if we are able to pre-generate a sufficiently diverse set of agent personas, we will be able to fit observed choices by a re-weighting of these decision makers. 

This mixture view unifies both settings: in the conjoint analysis, observed customer-level choices allow persona decisions to be corrected and aggregated prior to inference. In the regional experiment, where only regional shares are available, we explicitly learn aggregation weights $\theta_i$ to form region-level outcomes (Section~\ref{sec:mix-bias-corr}).

\section{Overall Framework}\label{sec:overall-framework}

This section lays out the general \textbf{Generation–Correction–Inference} LLM experiment pipeline, which anchors LLM-generated data to a small set of real human observations to support downstream economic inference.

\subsection{Generation–Correction–Inference}

Our framework aims to solve a fundamental causal inference challenge: estimating how outcomes ($Y$) respond to interventions ($T$) in different contexts ($x$). Technically, we seek to estimate $E[Y|T,X] = g(T,x)$, which under unconfoundedness has a causal interpretation as $E[Y(T)|x]$.

Our solution transforms this causal inference problem into quantitative estimates through three stages: First, in the \emph{Generation} stage, we leverage LLMs to create synthetic responses through different ``personas.'' Each persona $i$ provides counterfactual predictions $z_i=l_i(T,x)$. Next, in the \emph{Correction} stage, we use a limited human-labeled dataset $y$ to learn a mapping:
\[
f: (x, z) \;\longmapsto\; \hat y,
\]
thereby adjusting LLM outputs to better reflect actual human behavior. Finally, in the \emph{Inference} stage, we apply standard econometric or causal models on the corrected pairs $(x,\hat y)$ to estimate economic quantities $\theta$. A summary of the pipeline is shown in Table~\ref{tab:gci_pipeline}.

\begin{table*}[htbp]
  \centering
  \small
   \begin{tabular}{@{}p{3cm}p{5cm}p{5cm}@{}}
    \toprule
    & \textbf{Conjoint Analysis} & \textbf{Regional Experiment} \\ \midrule
    Decision granularity & \textit{customer}-level choice across profiles & \textit{regional}-level shares across the nation \\
    Calibration leverage & customer ground-truth labels & Aggregated region-level ground-truth \\
    Total observations & 4 millions+ & 35k+ region $\times$ day \\ 
    Synthetic scale & 1.2k synthetic customer choices & 17k synthetic personas \\ \bottomrule
  \end{tabular}
  \caption{Contrasting the two tasks and their value as a progression}
  \label{tab:task-comparison}
\end{table*}

\subsection{Two Use Cases}

To validate both the accuracy and scalability of our pipeline, we examine two use cases of increasing complexity. In a \emph{Conjoint Analysis} with millions of observations from tens of thousands of real customers, we model customer-level choices in response to order-level delivery-attribute changes. The bias correction model is trained only on 10\% fully-observed customer-level choices to generate the remaining 90\% choices synthetically. We then transfer the pipeline to a \emph{Regional Experiment} on delivery options, where treatments are assigned by ZIP3 and only aggregated regional delivery option shares are observed (906 ZIP3s $\times$ 60 days). We created 17k+ synthetic customer personas across regions to simulate order placements, and learned mixture weights to aggregate persona-level choices into calibrated regional outcomes. Table~\ref{tab:task-comparison} summarizes the comparison.

\section{Agentic Modeling for Conjoint Analysis}\label{sec:agent-conjoint}

\subsection{Problem Definition and Generation Stage}

Formally, let $c_{primary}=\{c_1,c_2,\ldots,c_M\}$ denote a set of primary customers. For each customer $c_i$, there is an associated profile vector $d_i\in\mathbb{R}^D$. Let $\mathcal{K}=\{1,2,3,\ldots,k\}$ denote the set of possible options for each task. Each customer $c_i$ completes a sequence of conjoint tasks represented by $s_i=[(x_{i1},y_{i1}),\ldots,(x_{it_i},y_{it_i})]$, where $x_{ij}\in\mathbb{R}^{qk}$ denotes the concatenated features of $k$ options in the $j^{th}$ question; $y_{ij}\in \mathcal{K}$ denotes the customer's response. Separately, an LLM is tasked with completing the same set of conjoint tasks, producing choices $z_{ij} \in \mathcal{K}$. An introduction to conjoint analysis is provided in Appendix~\ref{app:conjoint-intro}.

\subsection{Bias Correction for Conjoint Analysis}\label{sec:bias-correction-conjoint}

\begin{figure*}[htbp]
    \centering
    \includegraphics[width=\linewidth]{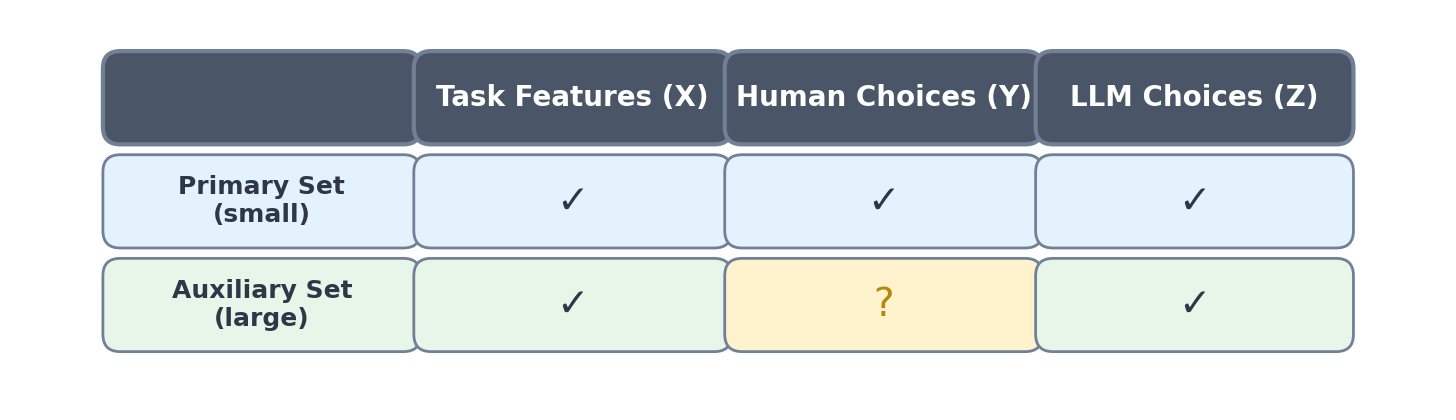}
    \caption{The general overview of the missing data situation. We assume we have task features, human choices, and LLM choices for the smaller primary set. We have task features and LLM choices for the larger auxiliary set. We don't have human choices for the auxiliary set.}
    \label{fig:conjoint-general-bias-corr}
\end{figure*}

Due to the high cost of conjoint analysis, we typically assume the size of the primary set is small. Another set of hypothetical auxiliary customers $c_{aux}$ ($N \gg M$), for whom actual human choice $\mathbf{y}$ is not collected. We can use LLMs to generate synthetic choice $\mathbf{z}$ for $c_{aux}$. However, prior work~\citep{wang2024large,goli2024frontiers} shows that relying directly on LLM choices for demand estimation is suboptimal. Thus, we debias LLMs by \textbf{using LLM choices $\mathbf{z}$ and option features $\mathbf{x}$ to predict counterfactual human choices $\mathbf{\hat{y}}$}. The missing-data situation is shown in Figure~\ref{fig:conjoint-general-bias-corr}.

To predict the human choices of $c_{aux}$, we use the primary customer data to learn the conditional distribution $\mathbb{P}(y\mid x,z)$ following~\citep{wang2024large}. The process is:
\begin{itemize}
    \item Fit a model $f$ to approximate $\mathbb{P}(y\mid x,z)$, using primary dataset $\mathcal{D}_{primary}$.
    \item Apply the fitted model $f$ on the auxiliary data to get a soft choice $\hat{y}_{aux}$ for auxiliary tasks.
\end{itemize}

\subsection{Inference: Demand Estimation}

One common practice for conjoint study is to estimate the part-worth $\beta \in \mathbb{R}^q$ for each attribute by fitting a Multinomial Logit (MNL) model. The best-in-class estimator minimizes:
\begin{multline}
\beta^* \in \arg\min_{\beta\in\mathbb{R}^d}\Biggl\{
\mathbb{E}_x\bigl[\mathrm{KL}(\mathbb{P}(y\mid x)\,\|\,\sigma_y(x;\beta))\bigr]\Biggr\}
\end{multline}
where $\sigma_j(x;\beta)=\frac{e^{x^{\top}_{(j)}\beta}}{1+\sum_{l\in\mathcal{K}}e^{x^{\top}_{(l)}}\beta}$.

\subsection{Experiment Setup and Evaluation}

We use both a general-purpose pre-trained LLM (Claude Sonnet 3.5v2) and an LLM fine-tuned (Qwen3-4B) for the conjoint task to generate synthetic choices $\mathbf{z}$. The bias correction model is a logistic regression model. \textbf{The notations of the part-worth coefficients estimated from different sets of choices} and other settings are in Appendix~\ref{app:mnl} and~\ref{app:con-data-model}. We measure the difference between the part-worth coefficients $\beta$ estimated from different sets of choices and the ground-truth part-worth coefficients $\beta^*$ using mean absolute percentage error (MAPE) and compare the change relative to MAPE($\beta^{primary}$) (Appendix ~\ref{app:con-eval-metric}). 

\subsection{Results and Analysis}
\begin{table}[htbp]
\centering
\resizebox{\columnwidth}{!}{%
\begin{tabular}{l|c|c|c|c}
\hline
\textbf{Estimators} & $\boldsymbol{\beta^{\text{aux}}}$ & $\boldsymbol{\beta^{\text{naive}}}$ & $\boldsymbol{\beta^{\text{tuned}}}$ & $\boldsymbol{\beta^{\text{pre-trained}}}$ \\
\hline
$\Delta$ MAPE($\boldsymbol{\beta}$) & 0.11 & -8.42 & -16.52 & -16.63 \\
\hline
\end{tabular}%
}
\caption{MAPE change from $\beta^{primary}$ for different estimators. \textbf{Negative} values indicate improvement over baseline. The definition of different estimators are in Appendix~\ref{app:mnl}.}
\label{tab:conjoint-results}
\end{table}

The bias reductions of different methods are shown in Table~\ref{tab:conjoint-results}. \textbf{Negative} values indicate improvement over baseline. Using predicted human choices from LLM generated choices via bias correction ($\beta^{pre-trained}$ and $\beta^{tuned}$) provides more accurate demand estimation. Directly using LLM generated choices ($\beta^{aux}$) even increases the bias, highlighting the necessity of bias correction. Alignment between LLM-generated and human choices exhibits substantial heterogeneity across customers (Appendix~\ref{app:prompt-conjoint}), further supporting the need for bias correction.

\begin{figure}[htbp]
    \centering
    \includegraphics[width=\linewidth]{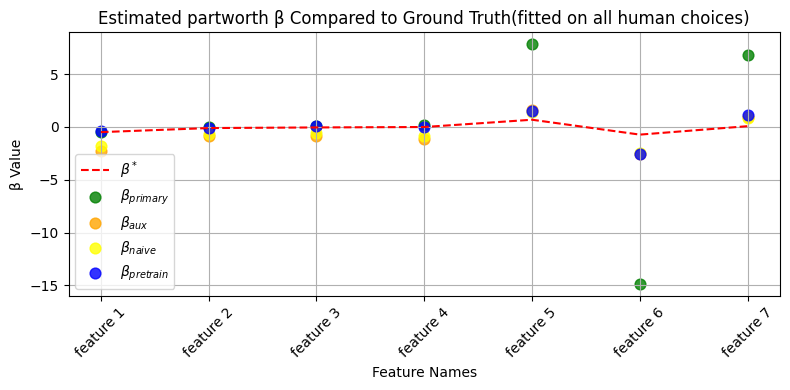}
    \caption{Partworth for different features in the conjoint task estimated based on different choice sets.}
    \label{fig:partworth}
\end{figure}

We visualize the coefficients of the part-worth in Figure~\ref{fig:partworth}. The part-worth estimated using bias correction based on LLM choices from Claude is closer to the ground-truth than other estimators. 

\section{Agentic Modeling for Regional Experiment}

\paragraph{Problem description}
The subsequent LLM simulation is based on a regional experiment. In this experiment, new features were added to the expedited delivery options for online shopping in the treatment group. A new feature was introduced for the Same-Day delivery option, while a different new feature was added to the Next-Day and Second-Day delivery options. The results showed that the share of Same-Day delivery decreased by 60 bps, while the share of Next-Day and Second-Day delivery increased by 25 bps. The Same-Day adjustment was associated with a more consistent directional effect, whereas the modifications to the Next-Day and Second-Day options exhibited negative or mixed effects according to other metrics.

Regional experiments pose greater challenges than conjoint tasks: they must run nationwide, are costly, and can cause negative effects (e.g., revenue loss). To mitigate this, we propose reducing scale by reducing the number of treatment regions and then using LLMs to simulate the remainder. The bias correction model must generalize to held-out regions whose human outcomes are never used in training. Unlike the conjoint setting, the regional experiment lacks customer-level choices; therefore, bias correction is performed via a mixture-of-personas model that learns aggregation weights to combine persona-level decisions into regional outcomes. We aim to examine whether, when an LLM replaces humans in conducting the experiment, the treatment effects \textbf{(1) are in the same direction as those measured in human experiments, (2) differ in magnitude (and if so, by how much), and (3) whether any such bias can be corrected}.

\subsection{Generation Stage}

We simulate regional demand responses by pairing region-specific personas with historically observed orders. For each ZIP3 area $z$, we build a persona set $P_z=\{p_1,\dots,p_{m_z}\}$ that captures local heterogeneity in preferences. Then, we sample a set of $n_z$ historical orders, denoted by $\mathcal{O}_z$. Each order is presented with a delivery menu that mirrors the original experiment: (i) an FST option (free shipping when the order exceeds \$35), (ii) a standard option (deliver strictly faster than FST but slower than the expedited option), (iii) an expedited option---same-, next-, or second-day---selected at random subject to ZIP3 availability, and (iv) a give-up (no-purchase) option. In the treatment group, expedited options were added with new features and the original configuration was kept in the control group. Acting under persona $p\in P_z$, the LLM chooses among the $K$ options for order $o\in\mathcal{O}_z$, yielding a one-hot vector $c^l\in\{0,1\}^K$. For a ZIP3 $z$, this generates triplets $\mathcal{T}_z=\{(o_i,p_i,c_i^l)\}_{i\le n_z*m_z}$, which serve as inputs to our downstream debiasing and inference pipeline. Prompts, persona, and order construction details are in Appendix~\ref{app:regional-details}.
\subsection{Bias Correction for Regional Experiment}\label{sec:mix-bias-corr}

Although the LLM often predicts the correct treatment direction, its effect size differs from human results, requiring bias correction similar to conjoint. Yet unlike conjoint tasks with order-level data, regional experiments lack customer-level choices, so loss must be computed at the \textbf{ZIP3 level}. 

Let $s_{z}^h\in \mathbb{R}^K$ denote the vector of delivery option shares for area $z$ observed in a small scale human experiment. As discussed in Section~\ref{sec:mix-model}, our mixture bias correction model learns a weight for each triplet $t_i=(o_i,p_i,c_i^l)$. Suppose the encoded feature representation of persona $p_i$ and order $o_i$ is $h_i$. The input to the model is a batch of such representations from the same ZIP3 $H=\{h_1,...h_N\}$, then, similar to the attention mechanism~\citep{bahdanau2014neural}, the weight for a pair of persona and order is:
\begin{equation}
    \theta_i = \frac{\exp(U^{\top}\tanh(Vh_i))}{\sum_{k=1}^N\exp(U^{\top}\tanh(Vh_k))}
\end{equation}
where $U$ and $V$ are trainable parameters. We compute the ZIP level share as $s_z^l=\sum_{i=1}^N\theta_i*c_i^l$ and minimize the KL divergence loss:
\begin{equation}
    \mathcal{L}_{KL}=\sum_z \sum_{j=1}^K s^{h}_{zj} \log\left(\frac{s^{h}_{zj}}{s^{l}_{zj}}\right)
\end{equation}

The input to the mixture model is a preprocessed numerical representation of triplet $(o_i, p_i)$. Order features $o_i$ include: product price, delivery option features, a product embedding generated by Sentence Transformer~\citep{reimers2019sentence}, and treatment-related features. Persona features $p_i$ consist of inferred demographics: age band, gender, income band, and education level. 

In addition to the mixture model, we also compare against an integrated model as an ablation of the correction architecture. Given a triplet $t_i$ as input, the model $f$ outputs a predicted human choice $g_i=\text{softmax}(f(t_i))$ where $g_i\in\mathbb{R}^K$. We aggregate all order-level predicted human choices within the same ZIP3 to obtain the ZIP3-level share $s_z^l=\frac{1}{|\mathcal{T}_z|}\sum_{t_i\in\mathcal{T}_z}g_i$. We also implement a prompting bias correction method by injecting prior delivery option shares into each LLM decision prompt. Details of the architecture of models and setup are in Appendix~\ref{app:model-arch} and~\ref{app:prompt-bias-correction}.

To reduce the scale of the experiment, we can conduct the human experiment in only a small subset of regions (ZIP3s), and then train the model on these regions, $\mathcal{Z}_{ID}$ (in-domain; the calibration set), to evaluate whether it can generalize to new regions, $\mathcal{Z}_{OOD}$ (out-of-domain; held out from training). The bias correction model here aims to learn human share based on the persona distribution and LLM choice in regions where human experiments have been conducted. The key assumption is that this learned relationship can generalize across variations in persona distributions, allowing us to extrapolate to regions where no human experiments were performed.

\subsection{Inference: Treatment Effect Estimation}

To quantify the impact of the treatment, we employ a difference-in-differences (DiD) research design:
\begin{multline}
    Y_{zt} = \beta_0 + \beta_1 \cdot Treatment_z + \beta_2 \cdot Post_t \\
    + \beta_3 \cdot (Treatment_z \times Post_t) + \epsilon_{zt}
\end{multline}
where $Y_{zt}$  represents the outcome variable for ZIP3 $z$ at time $t$ (the share of each delivery option), $Treatment_z$  is an indicator for treatment zip codes, $Post_t$ is an indicator for the post-treatment period, and $\epsilon_{zt}$ is the error term. The coefficient of interest is $\beta_3$, which captures the treatment effect.

\subsection{Experiment Setup}

We selected over 15k+ U.S.-based customers from a proprietary dataset to construct our personas. We followed the methodology outlined in~\citet{chen2024empathy} to generate personas based on customers' demographic information. In the region-wise setting, the sampling fraction $r$ for $\mathcal{Z}_{ID}$ is set to 0.1. 

We used a stratified sampling strategy to select $\mathcal{Z}_{ID}$. In the original regional experiment, there were 229 treatment ZIP3 areas with SSD launched and 220 without SSD. In the control group, there were 245 ZIP3 areas with SSD and 212 without. We randomly sample a fraction $r=0.1$ of ZIP3s from each of the four Treatment $\times$ SSD groups to form $\mathcal{Z}_{ID}$. The remaining ZIP3s are assigned to $\mathcal{Z}_{OOD}$. This ensures that $\mathcal{Z}_{ID}$ includes both treatment and control ZIP3s from the original experiment, as well as ZIP3s with and without SSD.

\subsection{Confidence Interval Construction}\label{app:conf-int}

When using LLM agents to conduct human experiments, the cost of invoking LLMs is relatively low compared to recruiting real human participants. This allows us to obtain a large number of LLM-generated samples. Applying methods such as DiD to estimate treatment effects would yield an infinitesimally small confidence interval. Therefore, in LLM-based experiments, one cannot rely on sampling variance alone to measure the confidence interval. Instead, it is necessary to account for the variance introduced throughout the entire statistical process, including generation, correction, and inference. 

Here, we adopt a bootstrap strategy: we repeat all experiments 20 times, each time randomly sampling with replacement from the set of personas and from regions in the region-wise experiments (which correspond to the generation stage), and retraining the bias correction model (correction stage). The treatment effect is then estimated based on the bias correction model’s new outputs (inference stage). This procedure yields 20 estimated treatment effects, from which we construct the 95\% confidence interval assuming these estimates follow a t-distribution. When synthetic outputs are abundant, the sampling variance of the synthetic 
sample becomes negligible, and the dominant uncertainty comes from the limited human data used for correction. Applying DiD to the corrected outputs alone would therefore understate the true uncertainty. We instead use a full-pipeline bootstrap---resampling personas and regions, retraining the correction model, and 
rerunning inference---so that the uncertainty induced by the finite human sample propagates through the estimate. This captures the variability of the corrector, but not its specification error.

\subsection{Results of Region-wise Bias Correction}\label{sec:region-wise-result}

\begin{table}[htbp]
\centering
\begin{tabular}{l|c}
\toprule
\textbf{Data Source} & \textbf{Effect (bps)} \\
\midrule
Human exp. in $\mathcal{Z}_{ID}$ only & $-57 \pm 27$ \\
Human exp. in $\mathcal{Z}_{OOD}$ only & $-60 \pm 9$ \\
Integrated model on $\mathcal{Z}_{OOD}$ & $-39 \pm 9$ \\
Mixture model on $\mathcal{Z}_{OOD}$ & $-65 \pm 10$ \\
National human experiments & $-60 \pm 8$ \\
\bottomrule
\end{tabular}
\caption{Region-wise results. CIs from 20 full-pipeline bootstrap replicates (resampling personas and regions with replacement and retraining the correction model). See Section~\ref{app:conf-int} for confidence interval details.}
\label{tab:region-wise-results}
\end{table}

The main purpose of the region-wise experiment is to evaluate the model's generalization ability. In Table~\ref{tab:region-wise-results}, the mixture model demonstrates stronger generalization ability (on $\mathcal{Z}_{OOD}$) than the integrated model. The estimated treatment effect by the mixture model (-65 $\pm$ 10 bps) is close to that of the complete human experiment (-60 $\pm$ 8 bps). The black-box integrated model produces only -39 $\pm$ 9 bps. The generalization ability demonstrated by the mixture model enables us to substantially \textbf{reduce the regions} required for conducting human experiments.

Taken together, the conjoint and regional experiments instantiate the same correction principle under different levels of observability. When individual choices are observed, correction is performed at the individual level by learning $P(y \mid x,z)$; when only aggregate outcomes are observed, correction must operate at the aggregation level. We think that AEM is most effective when raw LLM outputs are biased but still informative about human behavior. When raw LLM outputs contain limited signal about human outcomes, the correction step has less to exploit, and the gains from correction are naturally smaller. This is consistent with our observation that time-wise extrapolation (Appendix ~\ref{app:time-disadv}) is more challenging under the current pipeline.

\section{Conclusion}

This paper introduces Agentic Economic Modeling, a framework that leverages LLMs for bias-corrected counterfactual generation. Our Generation--Correction--Inference pipeline supports reliable inference via systematic bias correction against real human behavior. In a conjoint case study, the bias-corrected approach substantially reduces demand-estimation error using only 10\% of the survey data for the test product. The regional experiment demonstrates scalability, with the mixture model estimating treatment effects (-65$\pm$10 bps) closely matching the full human experiment (-60$\pm$8 bps). These results suggest that LLM-based economic modeling can complement traditional methods by reducing experimental cost and enabling parallel experiments across products and regions. Beyond the two real-world experiments in this paper, we plan to further test the method in AB test production settings to better understand its generalizability. Another promising extension is a design-based residual correction: since assignment is randomized and sampling rates are known, the Riesz representer is analytically available, giving approximate unbiasedness even under correction-model misspecification. We leave this to future work.

\section*{Ethical Considerations}
This work uses proprietary datasets for validation purposes only. All personally identifiable information is either anonymized or replaced with placeholders, and no real customer identities are exposed. The personas used in our experiments are synthetically generated and do not correspond to real individuals.

To simulate human behavior, our experiments use LLM-generated outputs, which may encode biases present in pretraining data and the underlying language models. Our framework then relies on a small amount of human data to calibrate these outputs and mitigate such biases.

\bibliography{sample-base,general}

\appendix


\clearpage
\section{Conjoint Analysis}\label{app:conjoint-intro}

Conjoint experiments reveal customer preferences by asking participants to evaluate or choose among products with varying attributes, showing which features matter most and the trade-offs they are willing to make. Large tech companies have used conjoint analysis to optimize feature configuration and attribute prioritization in product launches. However, conjoint studies are costly and slow: analyzing a single product typically requires hundreds of respondents, about thousands of dollars in marginal costs, and several weeks to complete—by which time results may already be outdated. To address this, we propose running experiments on only a small subset of participants (e.g., 10\% of the original scale) and using LLMs to simulate the remainder.

\subsection{Datasets and Foundation Models}\label{app:con-data-model}

We use the survey data from a proprietary dataset. This dataset includes massive-scale choice-based conjoint questions to collect preferences from hundreds of survey respondents over a few questions relating to thousands of representative products. In each question, respondents are asked to select a store from a set of options consisting of four online stores with different delivery configurations and one physical store. There are millions of observations in this dataset. We randomly select one testing product for initial evaluation and split the customers associated with the product into a primary customer set $c_{primary}$ (10\%) and an auxiliary customer set $c_{aux}$ (90 \%). We have tuned Qwen3-$x$B~\citep{yang2025qwen3}, where $x\in\{0.6,1.7,4,14\}$ is the size of the model. For all 1500 products in the dataset, we split out 200 products for testing purpose and the other products for tuning.

\subsection{MNL Model Details and Estimator Definitions}\label{app:mnl}

The best-in-class estimator of the MNL model is:
\begin{multline}
\beta^* \in
\arg\min_{\beta\in\mathbb{R}^d}\Biggl\{
\mathbb{E}_x\bigl[\mathrm{KL}(\mathbb{P}(y\mid x)\,\|\,\sigma_y(x;\beta))\bigr] \\
= \mathbb{E}_x\Bigl[\sum_{j\in\mathcal{K}^+}\mathbb{P}(y=j\mid x)
\log\frac{\mathbb{P}(y=j\mid x)}{\sigma_j(x;\beta)}\Bigr]
\Biggr\}
\end{multline}
where $\sigma_j(x;\beta)=\frac{e^{x^{\top}_{(j)}\beta}}{1+\sum_{l\in\mathcal{K}}e^{x^{\top}_{(l)}}\beta}, \forall j\in \mathcal{K}$ is the approximation to the true choice probabilities given by the estimator.

The estimated $\beta$ coefficient of an attribute captures how much that level increases or decreases the overall systematic utility driving respondents' choices. When different attributes are included, relative part-worths can be compared to quantify trade-offs between attributes, translating utility differences into interpretable measures of preference strength.

We compare the following estimators for part-worth coefficients:
\begin{itemize}
    \item $\beta^*$: Ground truth based on \textbf{all available human choices} $\{(x_{primary} \cup x_{aux}, y_{primary} \cup y_{aux})\}$. In real-world applications, $y_{aux}$ is unavailable; we use it only for evaluation.
    \item $\beta^{primary}$: Based on \textbf{primary customer choices only}: $\{(x_{primary}, y_{primary})\}$
    \item $\beta^{aux}$: Based on \textbf{raw LLM choices} from pre-trained Claude model: $\{(x_{aux}, z_{aux})\}$
    \item $\beta^{naive}$: Based on a \textbf{naive combination} of primary human choices and auxiliary LLM choices from pre-trained Claude model: $\{(x_{primary} \cup x_{aux}, y_{primary} \cup z_{aux})\}$
    \item $\beta^{pre-trained}$: Based on \textbf{predicted human choices} using bias correction: $\{(x_{aux}, \hat{y}_{aux})\}$. The raw LLM choices $z_{aux}$ are from the pre-trained Claude model.
     \item $\beta^{tuned}$: Based on \textbf{predicted human choices} using bias correction: $\{(x_{aux}, \hat{y}_{aux})\}$. The raw LLM choices $z_{aux}$ are from the fine-tuned Qwen3-4B model.
\end{itemize}

\subsection{Evaluation Metric}\label{app:con-eval-metric}
We use mean absolute percentage error (MAPE): $\text{MAPE}(\beta) = \frac{1}{q}\sum_{i=1}^q \frac{|\beta_i - \beta_i^*|}{|\beta_i^*|}$. Following~\citet{wang2024large}, we report the change in MAPE relative to $\text{MAPE}(\beta^{primary})$ as the benchmark. 
For example, if $MAPE(\beta^{corr})-MAPE(\beta^{primary})$
is a negative number, it suggests that the bias correction method outperforms the primary-data-only estimator. We refer to this change as bias reduction (from primary estimator).

\subsection{Prompting Experiments for Conjoint Analysis}\label{app:prompt-conjoint}

\begin{figure}[htbp]
    \centering
    \includegraphics[width=\linewidth]{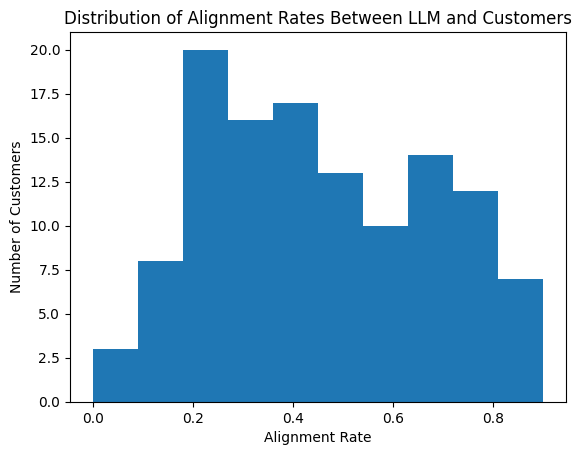}
    \caption{The distribution of alignment rates between the LLM's choices and each customer's choices. The alignment rates vary substantially across customers.}
    \label{fig:human-align-distribution}
\end{figure}

We first conducted experiments to see how well the model can perform the conjoint task based on different user preferences. We attempt to incorporate customers' profiles and historical choices into the prompt when prompting the LLM to make selections. We consider four variations:

\paragraph{Simple prompting the choice tasks directly} The input consists of only the textual description of the choice task and the LLM will generate choices $z$ based on that. The LLM-generated choices can be directly used to predict human choices and perform demand estimation.

\paragraph{Incorporating elementary customer profiles} Typically, before starting the actual conjoint choice tasks, customers are asked to provide demographic information, such as age, education level, membership status, and preferred delivering methods. We incorporate this information into the prompt, aiming for the LLM to make choices based on its understanding of the behaviors of these subgroups.

\paragraph{Incorporating historical choices of customers} For each customer $c_i$, we use the first $n$ completed tasks $s_i[:n]$ as historical choices, which are included in the prompt provided to the LLM to predict the remaining tasks $s_i[n+1:]$. Each remaining task in $s_i[n+1:]$ is treated independently; that is, the LLM responds to one task at a time given the historical choices. While subgroup-level behavioral preferences may provide some signal, they often fail to capture the full extent of customer heterogeneity. A customer's true preferences are more accurately reflected in their own historical choices.

\paragraph{Using a behavioral summary derived from historical choices} A potential drawback of incorporating customer historical choices is that the prompt length increases with the number of historical examples, which may negatively impact the LLM's reasoning performance~\citep{hsieh2024ruler}. To offer a stable-length alternative, given $s_i[:n]$ and $d_i$, we first ask the LLM to summarize the customer behavior $b_{in}$ in an independent run. We then incorporate $b_{in}$ into the prompt given to the LLM.

For this part of the analysis, the number of historical choices for each customer $n$ ranges from 1 to 8. To ensure a fair comparison across different numbers of historical choices, we fix the test set to be the tasks starting from the 9th task for each customer, regardless of how many historical choices are provided. For all prompting experiments, we use Sonnet-3.5v2.

The alignment rate without any profile information and historical choices is 42.8\%. This serves as a baseline for prompting-related methods. Figure~\ref{fig:human-align-distribution} shows the distribution of alignment rates between the LLM's choices and each individual customer's choices. The alignment rates vary substantially across individual customers and approximately follow a log-normal distribution. Assuming the LLM has its own default preferences and customers are rational, the figure suggests that LLM preferences differ from many of the customers and customers' preferences are diverse.

\begin{figure}[htbp]
    \centering
    \includegraphics[width=\linewidth]{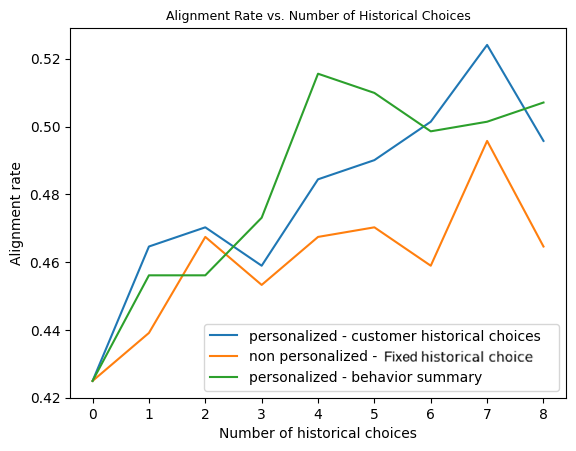}
    \caption{The performance of using customer historical choices, fixed (non-personalized) choices, and customer behavior summaries, with varying numbers of historical choices. The performance is generally better with more historical choices used. Personalized methods consistently perform better than non-personalized.}
    \label{fig:alignment-rate}
\end{figure}

Figure~\ref{fig:alignment-rate} shows the performance of different methods as the number of historical choices varies. A value of 0 indicates that only elementary profile information is incorporated. Overall, the alignment rate generally increases as more historical choices are added. When using the customer's own historical choices, an average alignment rate of 52.4\% is achieved with 7 examples (blue line). When using the generated behavior summary, the best alignment rate of 51.6\% is obtained when 4 historical choices are used to generate the summary (orange line). In comparison, the alignment rate using only elementary profile information is only 42.4\% ($n=0$).

When using a fixed set of example choices for all customers (non-personalized few-shot learning), increasing the number of choices improves performance. However, a performance gap persists relative to using customers' own historical choices, suggesting that personalized preferences contribute to the observed improvements.

\section{Regional Experiment}\label{app:regional-details}

\subsection{Regional Experiment Generation Stage Details}
\subsubsection{Persona Construction}\label{app:persona}

As mentioned earlier, LLMs are limited by their fixed customer preferences when using their outputs for demand analysis. To simulate customer heterogeneity, we use diverse personas representing varied customer preferences. This enables us to capture segment-level differences in LLM responses and improve the accuracy of experimental results. Due to differences in population distribution, the outcomes of such experiments are expected to vary significantly across regions. Ideally, for each ZIP3 area $z$, we aim to construct a region-specific persona set $P_z=\{p_1,p_2,...,p_{m_z}\}$ tailored to the demographic and behavioral characteristics of that area where $m_z$ is the total number of available personas of area $z$.

Our personas are derived from a proprietary dataset that records demographic details such as age, gender, income, zip code, and education level. We selected over 15k+ U.S.-based customers from the dataset to construct our personas. We followed the methodology outlined in~\citet{chen2024empathy} to generate personas based on customers' demographic information using Sonnet3.5v2. An example of a generated persona is as follows:

\begin{tcolorbox}[colback=gray!10, colframe=gray!80, fontupper=\small]
\{[Name]\} is a \{[X]\}-year-old \{[Race]\} \{[Gender]\} living at \{[Address]\}. She speaks English and her educational background includes some college credits but no degree from \{[College]\}. Her date of birth is \{[Date]\}. She is currently working as a part-time retail sales associate, with an annual income of \$\{[Price]\}. She is divorced and has three adult children. She primarily uses her smartphone to stay connected with her family through Facebook and text messages, and enjoys watching cooking videos on YouTube. On her laptop, she prefers using the built-in touchpad and frequently checks her email and online banking. She is somewhat cautious about internet security and usually asks her children for help with technical issues. She spends approximately 3 hours daily on her devices, mainly in the evenings after work, and uses her phone's voice-to-text feature frequently when messaging.
\end{tcolorbox}

The real customer information has been marked with placeholders. The rest of the information was generated by the model, which is not real.

\subsubsection{Order Construction}

The proprietary dataset includes a table containing the actual delivery options shown to customers at checkout. However, due to the table's massive size, it is not feasible for an LLM to conduct billions of experiments. Therefore, we sampled 3,806 placed orders $\mathcal{O}$ from the historical period of the experiment. (Note that these orders were not actually placed by the personas described above.) To simplify the experiment, we excluded all orders containing multiple products from the sample.

These orders come from ZIP3s across the country, with an average of 4.4 orders per ZIP3. The sampling of the set $\mathcal{O}_z$ ensures inclusion of all sampled orders that were actually placed in the given ZIP3. However, since each ZIP3 has only about 4.4 orders on average, the remaining of the 42 orders inevitably come from other regions. The number $n_z=42$ is chosen to simulate a 6-week LLM experiment.

The delivery options for each order were configured as follows: If the order value exceeded a specified threshold (e.g., \$35), a free-shipping FST option was included, with a delivery day sampled from the empirical FST distribution. A standard delivery option was then added, with a delivery time constrained to be shorter than the sampled FST time. An expedited option—same-day, next-day, or second-day delivery—was also included, selected at random subject to ZIP3 availability (if same-day service was unavailable, only next- or second-day options were used). In the treatment group, the expedited delivery options were configured with new features that differed from those in the control group, where the original configuration was retained. Finally, a give-up option was included, allowing the customer to choose not to place the order.

\subsubsection{Persona-Order Interaction and LLM Choice Generation}

The basic idea of using an LLM to conduct this randomized experiment is to prompt the LLM, acting as a given persona, to choose which delivery option they would select when placing a specific order.

For each ZIP3 $z$, we sampled $n_z$ orders $\mathcal{O}_z =\{o_1 ,o_2 ,...,o_{n_z} \}$ from the pool $\mathcal{O}$. Next, we have all personas in $P_z$ select a delivery option for each of the $n_z$ sampled orders in their corresponding ZIP3. Let $c_{ij}^l$ be the LLM choice when acting as $p_j \in P_z$ to place order $o_i \in \mathcal{O}_z$. Here, a choice $c\in\mathbb{R}^K$, where $K$ is the total number of available delivery options. A value of $c_k =1$ and $c_q =0$ for $q\neq k$ indicates that the LLM selected the $k^{th}$ option. For simplicity, the subscripts indicating the dependence of $o$, $p$ and $c$ on $z$ are omitted. This process generates a list of triplets
$\mathcal{T}_z = [(o_1, p_1, c_{11}^l), \ldots, (o_{n_z}, p_{m_z}, c_{n_z m_z}^l)]$ for each $z$. For simplicity, let $t_i =(o_i,p_i,c_i^l)$ be the $i^{th}$ triplet in $\mathcal{T}z$ for $1\leq i\leq n_z*m_z$. The prompt used to have the LLM make delivery choices is shown in Appendix \ref{app:reg-llm-prompt}, which is similar to the one used in the earlier conjoint experiment. In the experiments, $n_z$ is set to 42 for all ZIP3 areas. 

\subsection{Bias Correction Model Architecture}\label{app:model-arch}

\subsubsection{Mixture Model}

The concatenated features of $o_i$ and $p_i$ are passed through a linear layer with a hidden dimension of 64, which is also the input dimension of $U$. The hidden dimension of the attention layer (between $U$ and $V$) is 32. Models are optimized with Adam~\citep{kingma2017adammethodstochasticoptimization} at learning rate 1e-5.

\subsubsection{Integrated Model (Baseline)}

For a fair comparison, the integrated model also uses three linear layers, with ReLU activation between the first two layers and tanh activation between the last two layers. In addition to $o_i$ and $p_i$, the integrated model's input also includes the LLM's choice $c_i^l$.

\subsection{Regional Experiment Setup}
\subsubsection{Prompt Given to LLM to Make Delivery Option Choice}\label{app:reg-llm-prompt}

The confidential information has been replaced with placeholders.
The rest of the information was generated by the model, which is not real. As shown, each triplet—consisting of an order, a persona, and a delivery option—is specified within the prompt. In the experiments, there are 638,484 such prompts in total.

\begin{tcolorbox}[colback=gray!10, colframe=gray!80, fontupper=\small]
Suppose you want to buy a \{[Product]\}. The price is \$\{[Price]\}. We will ask you one question about buying the product with different delivery options. Consider the product itself, delivery options when making your choice. 
You should act like this person when purchasing:
\\\\
\{[Name]\} is a \{[X]\}-year-old \{[Race]\} \{[Gender]\} living at \{[Address]\}. She speaks English and Spanish, and her educational background includes completing one year at \{[College]\}, where she is pursuing a degree in \{[Major]\}. Her date of birth is \{[Date]\}. She is currently working as a junior software developer at a tech startup, with an annual income of \$\{[Salary]\}. She is single and has no children. She is highly tech-savvy, spending approximately 6–8 hours daily on her MacBook Pro for both work and personal use. She prefers using dark mode on all her devices and frequently uses productivity apps like Notion and Trello. On her iPhone 14, she actively engages with social media platforms, particularly Instagram and TikTok, where she follows tech influencers and programming tutorials.
\\\\
Imagine you had the following delivery options. Which one would you prefer? You can assume the shopping options are identical in all respects not listed here. 
\\A (Delivery Option 1): [Details are hidden]
\\B (Delivery Option 2): [Details are hidden]
\\C (Option 3): Prefer not to buy it. 
\\Which option would you choose as this person? Your response should be a single letter A, B, or C.
\end{tcolorbox}

\subsubsection{Prompting Bias Correction}\label{app:prompt-bias-correction}

We implemented a prompting-based bias correction method by injecting prior delivery option shares into the LLM prompt. As shown in Table~\ref{tab:treat-effect}, prompting-based correction (+4 bps) performs poorly compared to model-based correction (-38 to -41 bps). This is likely because one option accounts for 97\% of the choices, and prompting encourages the model to overwhelmingly favor this ``safe'' option. Here is an example prompt:

\begin{tcolorbox}[colback=gray!10, colframe=gray!80, fontupper=\small]
We calculated the prior distribution of delivery preferences based on historical data of customers placing orders within the same zip code. \\
This data includes all product categories and all orders, excluding cases where customers decided not to complete the purchase (``Prefer not to buy it'').\\
From this filtered dataset (i.e., only including orders that were actually completed), we computed the following distribution of delivery preferences:\\
- Same-Day delivery: X\%\\
- Next-Day or second-day delivery: Y\%\\
- Longer delivery time (e.g., 3+ days): Z\%\\
These percentages sum to 100\% and represent customer preferences conditional on purchasing (i.e., excluding those who chose not to buy).\\
We do not observe or estimate the share of customers who chose not to buy the item (the last option), so this ``give-up'' rate is unknown.\\
Please take this prior into account when making a decision about the delivery option, assuming the customer has some typical delivery preference behavior based on local historical data.
\end{tcolorbox}

This text was added before the choice question in the prompt shown in Appendix~\ref{app:reg-llm-prompt}.

\subsection{Share Prediction Results}\label{app:share-group}

\begin{table*}[htbp]
\centering
\begin{tabular}{l|cccc}
\toprule
 & \textbf{Before} & \textbf{Prompt} & \textbf{Integrated} & \textbf{Mixture} \\
 & \textbf{BC} & \textbf{BC} & \textbf{Model} & \textbf{Model} \\
\midrule
SSD, T & 0.2518 & 0.6794 & 0.1109 & \textbf{0.0518} \\
SSD, C & 0.6519 & 0.6817 & 0.1414 & \textbf{0.1254} \\
Non-SSD, T & 0.3021 & 0.5094 & \textbf{0.0928} & 0.1777 \\
Non-SSD, C & 0.1366 & 0.5102 & 0.0939 & \textbf{0.0768} \\
\bottomrule
\end{tabular}
\caption{Performance of share prediction under different bias correction strategies for SSD and Non-SSD (Treatment and Control groups). We use the first week's human experiment as the ground truth. The lower the number, the better the performance.}
\label{tab:bias-correction-share}
\end{table*}

We compare four groups of delivery option shares against corresponding ground-truth shares from human experiments: 1) Shares in ZIP3 areas where SSD was launched and assigned to the treatment group 2) Shares in ZIP3 areas where SSD was launched and assigned to the control group 3) Shares in ZIP3 areas where SSD was not launched and assigned to the treatment group 4) Shares in ZIP3 areas where SSD was not launched and assigned to the control group.

As shown in Table~\ref{tab:bias-correction-share}, using a trained model for bias correction substantially improves the accuracy of share prediction, while using prompts for bias correction increases the bias. The mixture model achieved the best predictions in three groups.

\begin{table}[htbp]
\centering
\begin{tabular}{l|c}
\toprule
\textbf{Method} & \textbf{Effect (bps)} \\
\midrule
Full Human Experiment & $-60$ \\
W1 Human Experiment & $-37$ \\
LLM w/o BC & $-350$ \\
Prompting BC & $+4$ \\
BC with Integrated Model & $-38$ \\
BC with Mixture Model & $-41$ \\
\bottomrule
\end{tabular}
\caption{Effect on the share of Same-Day delivery across correction methods, reported in basis points. BC = bias correction.}
\label{tab:treat-effect}
\end{table}

\subsection{Time-wise Bias Correction Exploration}\label{app:time-disadv}
Besides region-wise bias correction, we can also conduct a time-wise prediction. Instead of conducting a full human experiment lasting more than two months, we can run a shorter human experiment for just a few days. We assume that we have the first $x=7$ days (one week) of human data in the time-wise bias correction setting.

\subsubsection{Results of Time-wise Bias Correction}\label{sec:time-wise-result}

As shown in Table~\ref{tab:treat-effect}, using the LLM's raw choices without bias correction yields a treatment effect that is directionally consistent with the human experiment but differs substantially in magnitude. By training the model for bias 
correction using the first week of human experiment data, we can substantially reduce this gap. When using only the human data for the first day, the corrected estimate 
changes from the human-only estimate of $-17$ bps to $-24$ bps, closer to the 
full-experiment effect of $-60$ bps. We do not report a confidence interval in this 
setting: as noted below, the region-wise bootstrap has no time-wise analogue, so we 
cannot resample alternative first days of the experiment, and the sampling uncertainty 
in the human data cannot be quantified here. We therefore report the point estimate only.

\begin{figure}[htbp]
    \centering
    \includegraphics[width=\linewidth]{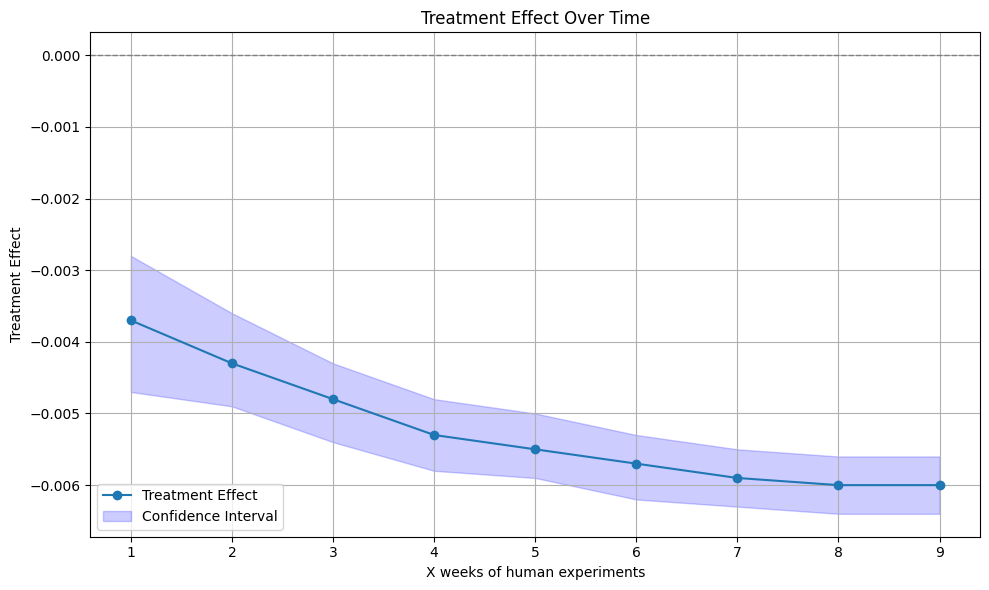}
    \caption{The treatment effect on the share of Same-Day delivery estimated using different numbers of weeks of human experiment data.}
    \label{fig:treat-effect-time}
\end{figure}
A gap to the full-experiment effect nonetheless remains, reflecting a limitation of the current pipeline: it does not model temporal change. As shown in Figure~\ref{fig:treat-effect-time}, as time progresses, the treatment effect tends to stabilize (though future changes cannot be ruled out), and the confidence interval narrows. The share of Same-Day delivery also fluctuates over time throughout the experiment (its change reached up to 40\% of the initial value). Since our LLM experiment does not incorporate temporal modeling, it cannot capture these temporal dynamics. In addition, unlike a region-wise experiment, a time-wise experiment does not allow for a similar bootstrap procedure. We cannot randomly sample different “first few days of the human experiments”, whereas in a region-wise experiment, we can always sample different regions to calculate the confidence interval.

Achieving temporal modeling in LLM experiments requires OOD prediction, since during LLM experimentation and bias correction, we cannot assume knowledge of such temporal trends. To make forward-looking predictions, we need to consider whether the changes in these variables are driven by people adapting to the effects of the experiment itself, or by external factors such as new users entering the system, holidays, or other contextual shifts. This is inherently a challenging task.

\end{document}